\documentclass[11pt,twoside]{article}
\usepackage{asp2010}

\resetcounters

\begin{document}

\title{Investigating stellar activity with CoRoT observations}
\author{S. Mathur$^{1}$,  D.~Salabert$^{3}$, R.A. Garc\'ia$^{2}$, C. R\'egulo$^{4,5}$, J. Ballot$^{6,7}$ and T.S.~Metcalfe$^{1}$\\
%{$^*$} Affiliations are given at the end of the paper}
\affil{$^1$High Altitude Observatory, NCAR, P.O. Box 3000, Boulder, CO 80307, USA}
\affil{$^2$Universit\'e de Nice Sophia-Antipolis, CNRS, Observatoire de la C\^ote dÕAzur, BP 4229, 06304 Nice Cedex 4, France}
\affil{$^3$Laboratoire AIM, CEA/DSM-CNRS-Universit\'e Paris Diderot; IRFU/SAp, Centre de Saclay, 91191 Gif-sur-Yvette Cedex, France}
\affil{$^4$Universidad de La Laguna, Dpto de Astrof\'isica, 38206, Tenerife, Spain}
\affil{$^5$Instituto de Astrof\'\i sica de Canarias, 38205, La Laguna, Tenerife, Spain}
\affil{$^6$CNRS, Institut de Recherche en Astrophysique et Plan\'etologie, 14 avenue Edouard Belin, 31400 Toulouse, France}
\affil{$^7$Universit\'e de Toulouse, UPS-OMP, IRAP, 31400 Toulouse, France}
}

\begin{abstract}
Recently, the study of the CoRoT target HD~49933 showed evidence of variability of its magnetic activity. This was the first time that a stellar activity was detected using asteroseismic data. For the Sun and HD~49933, we observe an increase of the p-mode frequencies and a decrease of the maximum amplitude per radial mode when the activity level is higher. Moreover a similar behavior Êof the frequency shifts with frequency has been found Êbetween the Sun and HD~49933.
We study 3 other targets of CoRoT as well, for which modes have been detected and well identified: HD~181420, HD~49385, and HD~52265 (which is hosting a planet). We show how the seismic parameters (frequency shifts and amplitude) vary during the observation of these stars. 
\end{abstract}

\section{Introduction}
Many important questions remain unanswered concerning the detailed mechanisms ruling the solar dynamo and activity cycle. The prediction on the length and strength of the solar activity cycles still needs additional constraints to be more reliable \citep{2008JApA...29...29D}. Different models of dynamos exist (interface dynamos, flux transport models including the $\alpha \Omega$ dynamo...).  All these models are based on the interaction between convection, rotation (in particular surface differential rotation), and magnetic field. One way to improve our understanding of this interaction is to study magnetic activity cycles in other solar-type stars providing different conditions in which dynamo can take place (rotation rates, depths of the convective zone...) \citep[e.g.][and references therein]{2008JPhCS.118a2032R}.

Many proxies are used to study magnetic activity cycles such as spectroscopic observations in CaHK, Lyman H$\alpha$, radio fluxes... In particular, the Mount Wilson HK project allowed to measure the photospheric and chromospheric activity in a very large sample of stars \citep[e.g.][]{1995ApJ...438..269B} suggesting that activity cycles features depend on the types of stars and their evolutionary stages. Beside, from this wide survey, it appeared that two different branches exist: the inactive and active branches \citep{2007ApJ...657..486B}. Possible explanations could be different positions of the dynamo shell or changes in the $\alpha$ effect. An empirical law has also been derived linking the surface rotation period of the star and the cycle period \citep{1997A&A...323..151O,2010A&A...509A..32J}. But the poor accuracy of this relation shows the need of more observational constraints.

It also seems that short cycles are more common than expected. An example is the star $i$~Hor, which has one of the shortest cycle period measures of 1.6~years \citep{2010ApJ...723L.213M}. $\tau$ Boo is an interesting case, as it hosts a giant close-in planet and for which \citet{2009MNRAS.398.1383F} report a estimate cycle period of $\sim$~2~years. With missions such as CoRoT \citep{2006cosp...36.3749B} and {\it Kepler} \citep{2010Sci...327..977B}, we can  expect to detect magnetic activity in some of their solar-type targets \citep[][]{2011Sci...332..213C} .

\section{Analyses and conclusions}

Seismology is a very powerful tool that allows us to retrieve information on the solar/stellar interior. We know that for the Sun, when the magnetic activity cycle of a star increases, two parameters of the acoustic modes vary: the frequencies of the modes increase while their amplitudes decrease. These two parameters vary in an anti-correlated manner. In addition, seismology has the asset of detecting any change in the magnetic activity inside a star even though no evidence at the surface of the star is visible \citep{2009A&A...504L...1S}.

\noindent As CoRoT provides a few months of continuous and very good quality of data, we analyzed the time series of one of the targets HD~49933, using the A2Z pipeline \citep{2010A&A...511A..46M}. We observed for the first time a magnetic activity using seismology, for another star than the Sun \citep{2010Sci...329.1032G}. Indeed, this cycle has several similarities with the solar one \citep{2011A&A...530A.127S}, suggesting a cycle period of at least 120~days. The CaHK observations done at Cerro Tololo Observatory in Chile confirms the cycle.

\noindent We then extended our study to three other CoRoT targets for which we had spectropolarimetric observations from NARVAL: HD~181420 \citep{2009A&A...506...51B}, HD~49385 \citep{2010A&A...515A..87D}, and HD~52265 \citep{2011A&A...530A..97B}.  These three stars present at least an anti-correlation between the temporal variation of the amplitude and the frequency shifts.The star HD~181420 seems to be in a stable phase while HD~49385 and HD~52265 show an anti-correlation between the frequency shifts and the amplitudes, suggesting a modest increase of magnetic activity \citep{2011JPhCS.271a2045M, mathur2011_act}.

\acknowledgements 
 The CoRoT has been developed and is operated by CNES, with contributions from Austria, Belgium, Brazil, ESA, Germany and Spain. NARVAL is a collaborative project funded by France (R\'egion Midi-Pyr\'en\'ees, CNRS, MENESR, Conseil G\'en\'eral des Hautes Pyr\'en\'ees) and the European Union (FEDER funds). DS and RAG acknowledges the support from CNES. RAG and SM acknowledge the support of the ``PNPS''. NCAR is supported by the National Science Foundation.\\

\bibliographystyle{asp2010} 
\bibliography{/Users/Savita/Documents/BIBLIO_sav.bib}

\begin{thebibliography}{}
\expandafter\ifx\csname natexlab\endcsname\relax\def\natexlab#1{#1}\fi
\expandafter\ifx\csname url\endcsname\relax
  \def\url#1{\texttt{#1}}\fi
\expandafter\ifx\csname urlprefix\endcsname\relax\def\urlprefix{URL }\fi
\providecommand{\eprint}[2][]{\url{#2}}

\bibitem[{{Baglin} et~al.(2006){Baglin}, {Auvergne}, {Boisnard}, {Lam-Trong},
  {Barge}, {Catala}, {Deleuil}, {Michel}, \& {Weiss}}]{2006cosp...36.3749B}
{Baglin}, A., {Auvergne}, M., {Boisnard}, L., {Lam-Trong}, T., {Barge}, P.,
  {Catala}, C., {Deleuil}, M., {Michel}, E., \& {Weiss}, W. 2006, in 36th
  COSPAR Scientific Assembly, vol.~36 of COSPAR, Plenary Meeting, 3749

\bibitem[{{Baliunas} et~al.(1995){Baliunas}, {Donahue}, {Soon}, {Horne},
  {Frazer}, {Woodard-Eklund}, {Bradford}, {Rao}, {Wilson}, {Zhang}, {Bennett},
  {Briggs}, {Carroll}, {Duncan}, {Figueroa}, {Lanning}, {Misch}, {Mueller},
  {Noyes}, {Poppe}, {Porter}, {Robinson}, {Russell}, {Shelton}, {Soyumer},
  {Vaughan}, \& {Whitney}}]{1995ApJ...438..269B}
{Baliunas}, S.~L., {Donahue}, R.~A., {Soon}, W.~H., {Horne}, J.~H., {Frazer},
  J., {Woodard-Eklund}, L., {Bradford}, M., {Rao}, L.~M., {Wilson}, O.~C.,
  {Zhang}, Q., {Bennett}, W., {Briggs}, J., {Carroll}, S.~M., {Duncan}, D.~K.,
  {Figueroa}, D., {Lanning}, H.~H., {Misch}, T., {Mueller}, J., {Noyes}, R.~W.,
  {Poppe}, D., {Porter}, A.~C., {Robinson}, C.~R., {Russell}, J., {Shelton},
  J.~C., {Soyumer}, T., {Vaughan}, A.~H., \& {Whitney}, J.~H. 1995, \apj, 438,
  269

\bibitem[{{Ballot} et~al.(2011){Ballot}, {Gizon}, {Samadi}, {Vauclair},
  {Benomar}, {Bruntt}, {Mosser}, {Stahn}, {Verner}, {Campante},
  {Garc{\'{\i}}a}, {Mathur}, {Salabert}, {Gaulme}, {R{\'e}gulo}, {Roxburgh},
  {Appourchaux}, {Baudin}, {Catala}, {Chaplin}, {Deheuvels}, {Michel}, {Bazot},
  {Creevey}, {Dolez}, {Elsworth}, {Sato}, {Vauclair}, {Auvergne}, \&
  {Baglin}}]{2011A&A...530A..97B}
{Ballot}, J., {Gizon}, L., {Samadi}, R., {Vauclair}, G., {Benomar}, O.,
  {Bruntt}, H., {Mosser}, B., {Stahn}, T., {Verner}, G.~A., {Campante}, T.~L.,
  {Garc{\'{\i}}a}, R.~A., {Mathur}, S., {Salabert}, D., {Gaulme}, P.,
  {R{\'e}gulo}, C., {Roxburgh}, I.~W., {Appourchaux}, T., {Baudin}, F.,
  {Catala}, C., {Chaplin}, W.~J., {Deheuvels}, S., {Michel}, E., {Bazot}, M.,
  {Creevey}, O., {Dolez}, N., {Elsworth}, Y., {Sato}, K.~H., {Vauclair}, S.,
  {Auvergne}, M., \& {Baglin}, A. 2011, \aap, 530, A97. \eprint{1105.3551}

\bibitem[{{Barban} et~al.(2009){Barban}, {Deheuvels}, {Baudin}, {Appourchaux},
  {Auvergne}, {Ballot}, {Boumier}, {Chaplin}, {Garc{\'{\i}}a}, {Gaulme},
  {Michel}, {Mosser}, {R{\'e}gulo}, {Roxburgh}, {Verner}, {Baglin}, {Catala},
  {Samadi}, {Bruntt}, {Elsworth}, \& {Mathur}}]{2009A&A...506...51B}
{Barban}, C., {Deheuvels}, S., {Baudin}, F., {Appourchaux}, T., {Auvergne}, M.,
  {Ballot}, J., {Boumier}, P., {Chaplin}, W.~J., {Garc{\'{\i}}a}, R.~A.,
  {Gaulme}, P., {Michel}, E., {Mosser}, B., {R{\'e}gulo}, C., {Roxburgh},
  I.~W., {Verner}, G., {Baglin}, A., {Catala}, C., {Samadi}, R., {Bruntt}, H.,
  {Elsworth}, Y., \& {Mathur}, S. 2009, \aap, 506, 51

\bibitem[{{B{\"o}hm-Vitense}(2007)}]{2007ApJ...657..486B}
{B{\"o}hm-Vitense}, E. 2007, \apj, 657, 486

\bibitem[{{Borucki} et~al.(2010){Borucki}, {Koch}, {Basri}, {Batalha}, {Brown},
  {Caldwell}, {Caldwell}, {Christensen-Dalsgaard}, {Cochran}, {DeVore},
  {Dunham}, {Dupree}, {Gautier}, {Geary}, {Gilliland}, {Gould}, {Howell},
  {Jenkins}, {Kondo}, {Latham}, {Marcy}, {Meibom}, {Kjeldsen}, {Lissauer},
  {Monet}, {Morrison}, {Sasselov}, {Tarter}, {Boss}, {Brownlee}, {Owen},
  {Buzasi}, {Charbonneau}, {Doyle}, {Fortney}, {Ford}, {Holman}, {Seager},
  {Steffen}, {Welsh}, {Rowe}, {Anderson}, {Buchhave}, {Ciardi}, {Walkowicz},
  {Sherry}, {Horch}, {Isaacson}, {Everett}, {Fischer}, {Torres}, {Johnson},
  {Endl}, {MacQueen}, {Bryson}, {Dotson}, {Haas}, {Kolodziejczak}, {Van Cleve},
  {Chandrasekaran}, {Twicken}, {Quintana}, {Clarke}, {Allen}, {Li}, {Wu},
  {Tenenbaum}, {Verner}, {Bruhweiler}, {Barnes}, \&
  {Prsa}}]{2010Sci...327..977B}
{Borucki}, W.~J., {Koch}, D., {Basri}, G., {Batalha}, N., {Brown}, T.,
  {Caldwell}, D., {Caldwell}, J., {Christensen-Dalsgaard}, J., {Cochran},
  W.~D., {DeVore}, E., {Dunham}, E.~W., {Dupree}, A.~K., {Gautier}, T.~N.,
  {Geary}, J.~C., {Gilliland}, R., {Gould}, A., {Howell}, S.~B., {Jenkins},
  J.~M., {Kondo}, Y., {Latham}, D.~W., {Marcy}, G.~W., {Meibom}, S.,
  {Kjeldsen}, H., {Lissauer}, J.~J., {Monet}, D.~G., {Morrison}, D.,
  {Sasselov}, D., {Tarter}, J., {Boss}, A., {Brownlee}, D., {Owen}, T.,
  {Buzasi}, D., {Charbonneau}, D., {Doyle}, L., {Fortney}, J., {Ford}, E.~B.,
  {Holman}, M.~J., {Seager}, S., {Steffen}, J.~H., {Welsh}, W.~F., {Rowe}, J.,
  {Anderson}, H., {Buchhave}, L., {Ciardi}, D., {Walkowicz}, L., {Sherry}, W.,
  {Horch}, E., {Isaacson}, H., {Everett}, M.~E., {Fischer}, D., {Torres}, G.,
  {Johnson}, J.~A., {Endl}, M., {MacQueen}, P., {Bryson}, S.~T., {Dotson}, J.,
  {Haas}, M., {Kolodziejczak}, J., {Van Cleve}, J., {Chandrasekaran}, H.,
  {Twicken}, J.~D., {Quintana}, E.~V., {Clarke}, B.~D., {Allen}, C., {Li}, J.,
  {Wu}, H., {Tenenbaum}, P., {Verner}, E., {Bruhweiler}, F., {Barnes}, J., \&
  {Prsa}, A. 2010, Science, 327, 977

\bibitem[{{Chaplin} et~al.(2011){Chaplin}, {Kjeldsen}, {Christensen-Dalsgaard},
  {Basu}, {Miglio}, {Appourchaux}, {Bedding}, {Elsworth}, {Garc{\'{\i}}a},
  {Gilliland}, {Girardi}, {Houdek}, {Karoff}, {Kawaler}, {Metcalfe},
  {Molenda-{\.Z}akowicz}, {Monteiro}, {Thompson}, {Verner}, {Ballot},
  {Bonanno}, {Brand{\~a}o}, {Broomhall}, {Bruntt}, {Campante}, {Corsaro},
  {Creevey}, {Do{\u g}an}, {Esch}, {Gai}, {Gaulme}, {Hale}, {Handberg},
  {Hekker}, {Huber}, {Jim{\'e}nez}, {Mathur}, {Mazumdar}, {Mosser}, {New},
  {Pinsonneault}, {Pricopi}, {Quirion}, {R{\'e}gulo}, {Salabert}, {Serenelli},
  {Aguirre}, {Sousa}, {Stello}, {Stevens}, {Suran}, {Uytterhoeven}, {White},
  {Borucki}, {Brown}, {Jenkins}, {Kinemuchi}, {Van Cleve}, \&
  {Klaus}}]{2011Sci...332..213C}
{Chaplin}, W.~J., {Kjeldsen}, H., {Christensen-Dalsgaard}, J., {Basu}, S.,
  {Miglio}, A., {Appourchaux}, T., {Bedding}, T.~R., {Elsworth}, Y.,
  {Garc{\'{\i}}a}, R.~A., {Gilliland}, R.~L., {Girardi}, L., {Houdek}, G.,
  {Karoff}, C., {Kawaler}, S.~D., {Metcalfe}, T.~S., {Molenda-{\.Z}akowicz},
  J., {Monteiro}, M.~J.~P.~F.~G., {Thompson}, M.~J., {Verner}, G.~A., {Ballot},
  J., {Bonanno}, A., {Brand{\~a}o}, I.~M., {Broomhall}, A., {Bruntt}, H.,
  {Campante}, T.~L., {Corsaro}, E., {Creevey}, O.~L., {Do{\u g}an}, G., {Esch},
  L., {Gai}, N., {Gaulme}, P., {Hale}, S.~J., {Handberg}, R., {Hekker}, S.,
  {Huber}, D., {Jim{\'e}nez}, A., {Mathur}, S., {Mazumdar}, A., {Mosser}, B.,
  {New}, R., {Pinsonneault}, M.~H., {Pricopi}, D., {Quirion}, P., {R{\'e}gulo},
  C., {Salabert}, D., {Serenelli}, A.~M., {Aguirre}, V.~S., {Sousa}, S.~G.,
  {Stello}, D., {Stevens}, I.~R., {Suran}, M.~D., {Uytterhoeven}, K., {White},
  T.~R., {Borucki}, W.~J., {Brown}, T.~M., {Jenkins}, J.~M., {Kinemuchi}, K.,
  {Van Cleve}, J., \& {Klaus}, T.~C. 2011, Science, 332, 213

\bibitem[{{Deheuvels} et~al.(2010){Deheuvels}, {Bruntt}, {Michel}, {Barban},
  {Verner}, {R{\'e}gulo}, {Mosser}, {Mathur}, {Gaulme}, {Garcia}, {Boumier},
  {Appourchaux}, {Samadi}, {Catala}, {Baudin}, {Baglin}, {Auvergne},
  {Roxburgh}, \& {P{\'e}rez Hern{\'a}ndez}}]{2010A&A...515A..87D}
{Deheuvels}, S., {Bruntt}, H., {Michel}, E., {Barban}, C., {Verner}, G.,
  {R{\'e}gulo}, C., {Mosser}, B., {Mathur}, S., {Gaulme}, P., {Garcia}, R.~A.,
  {Boumier}, P., {Appourchaux}, T., {Samadi}, R., {Catala}, C., {Baudin}, F.,
  {Baglin}, A., {Auvergne}, M., {Roxburgh}, I.~W., \& {P{\'e}rez
  Hern{\'a}ndez}, F. 2010, \aap, 515, A87. \eprint{1003.4368}

\bibitem[{{Dikpati} \& {Gilman}(2008)}]{2008JApA...29...29D}
{Dikpati}, M., \& {Gilman}, P.~A. 2008, Journal of Astrophysics and Astronomy,
  29, 29

\bibitem[{{Fares} et~al.(2009){Fares}, {Donati}, {Moutou}, {Bohlender},
  {Catala}, {Deleuil}, {Shkolnik}, {Collier Cameron}, {Jardine}, \&
  {Walker}}]{2009MNRAS.398.1383F}
{Fares}, R., {Donati}, J.-F., {Moutou}, C., {Bohlender}, D., {Catala}, C.,
  {Deleuil}, M., {Shkolnik}, E., {Collier Cameron}, A., {Jardine}, M.~M., \&
  {Walker}, G.~A.~H. 2009, \mnras, 398, 1383. \eprint{0906.4515}

\bibitem[{{Garc{\'{\i}}a} et~al.(2010){Garc{\'{\i}}a}, {Mathur}, {Salabert},
  {Ballot}, {R{\'e}gulo}, {Metcalfe}, \& {Baglin}}]{2010Sci...329.1032G}
{Garc{\'{\i}}a}, R.~A., {Mathur}, S., {Salabert}, D., {Ballot}, J.,
  {R{\'e}gulo}, C., {Metcalfe}, T.~S., \& {Baglin}, A. 2010, Science, 329,
  1032. \eprint{1008.4399}

\bibitem[{{Jouve} et~al.(2010){Jouve}, {Brown}, \&
  {Brun}}]{2010A&A...509A..32J}
{Jouve}, L., {Brown}, B.~P., \& {Brun}, A.~S. 2010, \aap, 509, A32.
  \eprint{0911.1947}

\bibitem[{{Mathur} et~al.(2010){Mathur}, {Garc{\'{\i}}a}, {R{\'e}gulo},
  {Creevey}, {Ballot}, {Salabert}, {Arentoft}, {Quirion}, {Chaplin}, \&
  {Kjeldsen}}]{2010A&A...511A..46M}
{Mathur}, S., {Garc{\'{\i}}a}, R.~A., {R{\'e}gulo}, C., {Creevey}, O.~L.,
  {Ballot}, J., {Salabert}, D., {Arentoft}, T., {Quirion}, P., {Chaplin},
  W.~J., \& {Kjeldsen}, H. 2010, \aap, 511, A46. \eprint{0912.3367}

\bibitem[{{Mathur} et~al.(2011{\natexlab{a}}){Mathur}, {Garc{\'{\i}}a},
  {Salabert}, {Ballot}, {R{\'e}gulo}, {Metcalfe}, \&
  {Baglin}}]{2011JPhCS.271a2045M}
{Mathur}, S., {Garc{\'{\i}}a}, R.~A., {Salabert}, D., {Ballot}, J.,
  {R{\'e}gulo}, C., {Metcalfe}, T.~S., \& {Baglin}, A. 2011{\natexlab{a}},
  Journal of Physics Conference Series, 271, 012045. \eprint{1011.4102}

\bibitem[{{Mathur} et~al.(2011{\natexlab{b}}){Mathur}, {Garc\'\i a},
  {Morgenthaler}, {Salabert}, {Petit}, {Ballot}, {R\'egulo}, \&
  {Catala}}]{mathur2011_act}
{Mathur}, S., {Garc\'\i a}, R.~A., {Morgenthaler}, A., {Salabert}, D., {Petit},
  P., {Ballot}, J., {R\'egulo}, C., \& {Catala}, C. 2011{\natexlab{b}}, \aap,
  submitted

\bibitem[{{Metcalfe} et~al.(2010){Metcalfe}, {Basu}, {Henry}, {Soderblom},
  {Judge}, {Kn{\"o}lker}, {Mathur}, \& {Rempel}}]{2010ApJ...723L.213M}
{Metcalfe}, T.~S., {Basu}, S., {Henry}, T.~J., {Soderblom}, D.~R., {Judge},
  P.~G., {Kn{\"o}lker}, M., {Mathur}, S., \& {Rempel}, M. 2010, \apjl, 723,
  L213. \eprint{1009.5399}

\bibitem[{{Ossendrijver}(1997)}]{1997A&A...323..151O}
{Ossendrijver}, A.~J.~H. 1997, \aap, 323, 151

\bibitem[{{Rempel}(2008)}]{2008JPhCS.118a2032R}
{Rempel}, M. 2008, Journal of Physics Conference Series, 118, 012032

\bibitem[{{Salabert} et~al.(2009){Salabert}, {Garc{\'{\i}}a}, {Pall{\'e}}, \&
  {Jim{\'e}nez-Reyes}}]{2009A&A...504L...1S}
{Salabert}, D., {Garc{\'{\i}}a}, R.~A., {Pall{\'e}}, P.~L., \&
  {Jim{\'e}nez-Reyes}, S.~J. 2009, \aap, 504, L1. \eprint{0907.3888}

\bibitem[{{Salabert} et~al.(2011){Salabert}, {R{\'e}gulo}, {Ballot},
  {Garc{\'{\i}}a}, \& {Mathur}}]{2011A&A...530A.127S}
{Salabert}, D., {R{\'e}gulo}, C., {Ballot}, J., {Garc{\'{\i}}a}, R.~A., \&
  {Mathur}, S. 2011, \aap, 530, A127. \eprint{1104.5654}

\end{thebibliography}

\end{document}